\documentclass{article}

\usepackage{spconf,amsmath,graphicx,multirow}
\usepackage{enumitem}
\usepackage{comment}
\usepackage{booktabs}
\usepackage{color, colortbl}
\usepackage[usenames,dvipsnames]{xcolor}
\usepackage{makecell}
\usepackage[normalem]{ulem}
\usepackage{adjustbox}
\usepackage{amssymb}
\usepackage{pifont}
\newcommand{\cmark}{\ding{51}}%
\newcommand{\xmark}{\ding{55}}%

\usepackage[labelformat=simple]{subcaption}

\usepackage[font=small,labelfont=bf]{caption}

\usepackage{url}
\usepackage[T1]{fontenc}

\renewenvironment{thebibliography}[1]{
  \begin{oldthebibliography}{#1}
    \setlength{\itemsep}{0em}
    \setlength{\parskip}{0em}
}
{
  \end{oldthebibliography}
}
\usepackage{setspace}

\newcommand{\COMMENT}[1]{}

\title{Training early-exit architectures for automatic speech recognition: fine-tuning pre-trained models or training from scratch}
\name{\begin{tabular}{c}George August Wright$^1$, Umberto Cappellazzo$^1$, Salah Zaiem$^2$, Desh Raj$^3$, \\Lucas Ondel Yang$^4$, Daniele Falavigna$^5$, Mohamed Nabih Ali$^5$, Alessio Brutti$^5$\end{tabular}}
\address{$^1$University of Trento; $^2$LTCI, Télécom Paris, Institut Polytechnique de Paris;\\
$^3$Johns Hopkins University; $^4$Universite Paris-Saclay, LISN, CNRS; $^5$Fondazione Bruno Kessler}

\begin{document}
\ninept
\maketitle

\begin{abstract} 
The ability to dynamically adjust the computational load of neural models during inference is crucial for on-device processing scenarios characterised by limited and time-varying computational resources. A promising solution is presented by early-exit architectures, in which additional exit branches are appended to intermediate layers of the encoder. In self-attention models for automatic speech recognition (ASR), early-exit architectures enable the development of dynamic models capable of adapting their size and architecture to varying levels of computational resources and ASR performance demands. Previous research on early-exiting ASR models has relied on pre-trained self-supervised models, fine-tuned with an early-exit loss. In this paper, we undertake an experimental comparison between fine-tuning pre-trained backbones and training models from scratch with the early-exiting objective. Experiments conducted on public datasets reveal that early-exit models trained from scratch not only preserve performance when using fewer encoder layers but also exhibit enhanced task accuracy compared to single-exit or pre-trained models. Furthermore, we explore an exit selection strategy grounded in posterior probabilities as an alternative to the conventional frame-based entropy approach. Results provide insights into the training dynamics of early-exit architectures for ASR models, particularly the efficacy of training strategies and exit selection methods.

\end{abstract}
\section{Introduction}
\label{sec:intro}

The edge-cloud continuum is an emerging complex ecosystem that integrates compute-enabled edge devices, distributing the overall computation workload among them~\cite{distributed}. The availability of computational resources differs considerably across devices, and varies over time due to resource sharing between multiple services. 
Therefore, it is crucial to have neural models that can dynamically change their trade-off between computation and performance. To this end, we investigate the use of early-exit architectures applied to large-vocabulary automatic speech recognition (ASR).

\begin{figure}[t]
  \centering
  \includegraphics[width=0.8\linewidth]{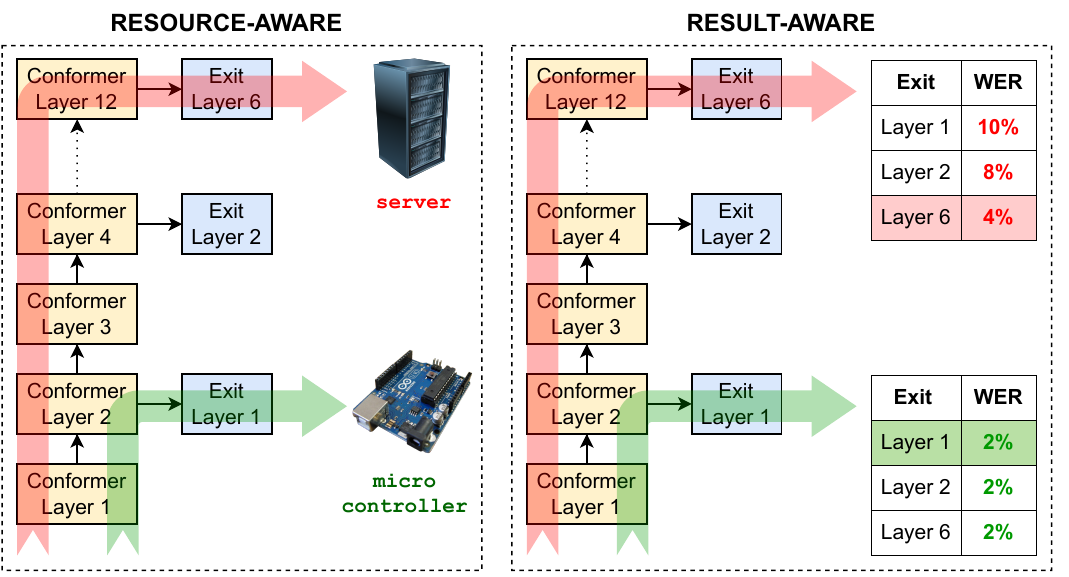}
\caption{Examples of \textit{resource-aware} and \textit{result-aware} uses of early-exits. Left: the micro-controller can afford only two layers, whereas the server can handle the entire model. Right: the first input requires processing with the entire network; in the second case, the best transcription is produced after only two encoder layers.}
\label{fig:awareness}
\end{figure}

Previous work for on-device neural processing focused on reducing model size through compression~\cite{cerutti2020}, knowledge distillation~\cite{gou2021knowledge, yu2013}, pruning~\cite{han2016}, and quantization~\cite{wu2016}. 
Although effective, these approaches provide static solutions which require handling multiple models with varying trade-offs. A preferable approach is to dynamically adapt a single model architecture to the memory and computational capabilities of each hosting device.
A solution is represented by ``early-exit'' architectures that introduce intermediate exit branches~\cite{branchynet2017, phuong2019}. 
The input is not processed by all layers of the neural network but only a subset of them, returning the result at an intermediate layer and bypassing operations in the layers not traversed. 
Figure~\ref{fig:awareness} displays examples of early-exiting networks, where layer-specific classifiers/decoders (the ``exit layers'') are appended to intermediate encoder layers. 
Early-exit architectures leverage the observation that, for simpler inputs, the lower layers of the model may have already learned sufficient parameters for accurate predictions. Early-exit architectures enable {\it resource-aware} processing (Fig.~\ref{fig:awareness}, left), where a single model can be used on heterogeneous devices, as well as {\it result-aware} processing (Fig.~\ref{fig:awareness}, right), where the model selects the earliest exit that achieves the same performance as the entire network.

\begin{figure}[ht]
    \centering
    \includegraphics[width=\linewidth,trim={0 0 6em 0},clip]{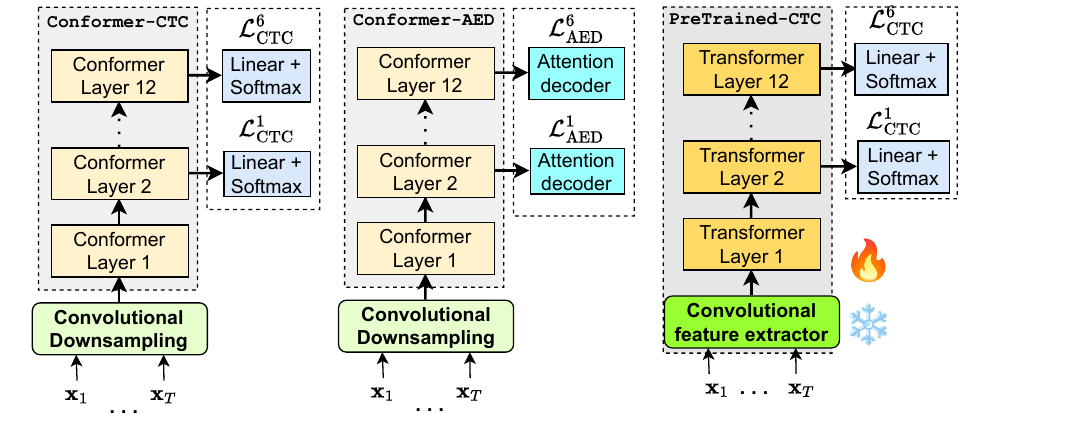}
    \captionof{figure}{Early-exit architectures (from left to right): Conformer-CTC, Conformer-AED, and pre-trained-CTC. Conformer-based models are trained from scratch. pre-trained-CTC are initialised from pre-trained models (Wav2Vec2.0 or WavLM) and fine-tuned with early-exit CTC losses, freezing the convolutional feature extractor.
    }
    \label{fig:arch-introduction}
\end{figure}

\begin{table}[ht]
    \centering
    \caption{Hyperparameters for the early-exit model architectures shown in Fig.~\ref{fig:arch-introduction}.}
    \setlength{\tabcolsep}{2pt}
    \adjustbox{max width=\linewidth}{
    \begin{tabular}{@{}lrrrr@{}}
    \toprule
    \textbf{Feature} & \begin{tabular}{r}\textbf{Conformer}\\\textbf{CTC}\end{tabular} & \begin{tabular}{r}\textbf{Conformer}\\\textbf{AED}\end{tabular} & \begin{tabular}{r}\textbf{Wav2Vec2}\\\textbf{CTC}\end{tabular}& \begin{tabular}{r}\textbf{WavLM (Base+)}\\\textbf{CTC}\end{tabular} \\
    \midrule
    \# params (M) & 31.0 & 13.3 & 94.0 &94.7\\
    Encoder & 12-layer Conf. & 12-layer Conf. &  12-layer Transf. & 12-layer Transf.\\
    Attention dim. & 256&144 & 768  & 768\\
    Number heads &8 & 4 & 8 & 8\\
    Feed-forward dim. & 2048& 1024 & 3072 & 3072\\
    Decoder & Linear & 4-layer Transf. & Linear & Linear\\
    Inputs & 80-d MFCC & 80-d MFCC & Waveform *  & Waveform * \\
    Loss function & $\mathcal{L}_{\mathrm{CTC}}$ & $\mathcal{L}_{\mathrm{CE}}$ \& $\mathcal{L}_{\mathrm{CTC}}$ & $\mathcal{L}_{\mathrm{CTC}}$  & $\mathcal{L}_{\mathrm{CTC}}$\\
    Output units & BPE & BPE & Grapheme & Grapheme\\
    LM rescoring & \xmark & \cmark & \xmark& \xmark \\
    Data augmentation & \xmark &\cmark & \cmark& \cmark \\
    \bottomrule
    \end{tabular}}
    \label{tab:hyperparams}
\end{table}%


In this work, we investigate the use of early-exit self-attention architectures applied to ASR models. 
Following common practice in ASR, previous work mainly focuses on accelerating inference using pre-trained backbones fine-tuned with early-exit objectives~\cite{hubertee2022, berrebbi2022avoid, zaiem2023}. In this paper we additionally investigate training models from scratch using different early-exit losses as depicted in Fig.~\ref{fig:arch-introduction}. We demonstrate that early-exit models, optimised on the joint loss of all early-exit layers, outperform ``single-exit'' models optimised only on the final layer output. Moreover, our experiments suggests that training from scratch native early-exit architectures is to some extent more effective than fine-tuning pre-trained self-supervised models. Our overall contributions are:

\begin{enumerate}[leftmargin=*,noitemsep]
\item We investigate early-exit training strategies, both training from scratch and fine-tuning pre-trained self-supervised models, considering 4 different architectures;
\item We evaluate a confidence-based exit selection based on N-best posteriors that offers a slightly better trade-off than entropy.

\end{enumerate}

\section{Related Work}
\label{sec:related}
Early-exiting methods were first introduced for computer vision in BranchyNet \cite{branchynet2017} by adding two branches to AlexNet \cite{alexnet2017}. 
The authors optimised the joint loss of the exits and defined a confidence measure, based on the entropy of the output class distribution, to decide the exit level. 
More recently, Scardapane et al.~\cite{scardapane2020} have provided a theoretical framework for multi-exit neural architectures.
Early-exit classifiers have also been implemented in tiny (KB-sized) models~\cite{ghanathe2023}. 
Besides early-exiting, other methods for dynamically adapting the model architecture for efficient inference have been explored, such as HydraNet~\cite{Mullapudi2018HydraNetsSD}.

In speech recognition, early-exiting was first introduced in HuBERT-EE~\cite{hubertee2022} to accelerate inference for a pre-trained HuBERT~\cite{hubert2021} model, using confidence measures based on CTC confidence or output entropies, 
with no significant performance degradation. Similarly, Zaiem et al.~\cite{zaiem2023} have investigated early-exit fine-tuning strategies in the context of a large pre-trained WavLM~\cite{wavlm2022} model, comparing them with approaches based on layer removal and input down-sampling.
The issue of overthinking in ASR encoders has been analysed in \cite{berrebbi2022avoid}, in which the authors report theoretical lower bounds of speed/quality trade-offs for early-exit strategies. Exit-selection strategies were proposed based on comparisons of output distribution and transcriptions between successive exits. Similar investigations using the entropy of the output distribution have also been conducted for recurrent neural networks~\cite{xin2022}. 

All aforementioned studies~\cite{hubertee2022, berrebbi2022avoid, zaiem2023} employ pre-trained models by fine-tuning the Transformer component, as is common for ASR. They are primarily focused on \textit{improving inference efficiency} by selecting the best early exit according to some criteria. Analogously, in natural language processing, research on early-exit strategies has focused on accelerating the inference of large pre-trained language models such as BERT~\cite{Zhou2020, Liu2020, Xin2020}. Conversely, in this work, our objective is to \textit{understand the training dynamics} of early-exit models (both trained from scratch and initialised from pre-trained models) by conducting exhaustive experiments on multiple datasets. We demonstrate that training the model from scratch,
jointly optimizing all exited layers of the model, provides significant performance improvements over conventional single-exit models and fine-tuned pre-trained models (particularly at the lowest exits).

\section{Early-Exit Models for ASR}
\label{sec:arch}

Given an input sequence $\mathbf{X}$, such as a raw waveform $\{x_1,\ldots,x_N\}$ or acoustic features $\{\mathbf{x}_1,\ldots,\mathbf{x}_T\}$, where $\mathbf{x}_t \in \mathbb{R}^d$, an ASR system estimates the output sequence $\hat{\mathbf{y}}$ as
\begin{equation}
\label{eq:asr}
\hat{\mathbf{y}} = \mathrm{arg}\max_{\mathbf{y}} P(\mathbf{y}|\mathbf{X}),
\end{equation}
where $\mathbf{y} \in \mathcal{Y}^{\ast}$, for some vocabulary $\mathcal{Y}$, such as graphemes, phonemes, or byte pair encoding (BPE) units. 
The distribution $P(\mathbf{y}|\mathbf{X})$ is usually estimated using a parameterised model $\Theta$ (such as a neural network), i.e., $P(\mathbf{y}|\mathbf{X};\Theta)$, which is learned using input-output pairs ($\mathbf{X}$,$\mathbf{y}$). 

For convenience, $\Theta$ is often factored into an encoder, which extracts high-dimensional representations $\mathbf{h}_1^T$ from $\mathbf{X}$, and a decoder, which maps $\mathbf{h}_1^T$ to the output sequence $\mathbf{y}_1^U$.
Since $U\ll T$ in general, ASR decoders either use (i) an \textit{alignment function} ($\mathcal{B}:\mathbf{a}_1^T\rightarrow \mathbf{y}_1^U$) for sequence training, or (ii) an \textit{attention mechanism} with label-based cross-entropy training.
We apply early-exiting to ASR by adding decoders at several intermediate layers of the encoder (as shown in Fig.~\ref{fig:arch-introduction}). Assuming that $M$ such intermediate exits are added (with hypothesis $\hat{\mathbf{y}}^1,\ldots,\hat{\mathbf{y}}^M$), the overall model is trained by optimizing the joint objective
\begin{equation}
\label{eq:jointloss}
    \mathcal{L}_{EE}(\hat{\mathbf{y}}^1,\ldots,\hat{\mathbf{y}}^M,\mathbf{y}) = \sum_{m=1}^{M}\mathcal{L}(\hat{\mathbf{y}}^m,\mathbf{y}),
\end{equation}
where $\mathcal{L}(\hat{\mathbf{y}}^m,\mathbf{y}) = -\log P(\mathbf{y}|\mathbf{X};\Theta_m)$, and $\Theta_m$ denotes the subset of $\Theta$ used for exit $m$.
In this work, we implement early-exiting in four diverse model architectures. Conformer-CTC and Conformer-AED share a Conformer encoder~\cite{conformerlu2019} architecture, with early exits appended to every other layer of a 12-layer encoder, but features different decoders. In addition we consider two pre-trained backbones based on Wav2Vec-2.0~\cite{baevski2020wav2vec2} and WavLM~\cite{wavlm2022} models.
Their hyperparameters are summarised in Tab.~\ref{tab:hyperparams}.

{\bf Conformer-CTC:} The Conformer encoder is used to obtain $\mathbf{h}_1^T$, and the decoder is a linear layer with softmax. The intermediate loss function is connectionist temporal classification (CTC)~\cite{Graves2006ConnectionistTC}:
\begin{equation}
\label{eq:ctc}
    \mathcal{L}_{\mathrm{CTC}}(\hat{\mathbf{y}},\mathbf{y}) = -\log \sum_{\mathbf{a}_1^T\in \mathcal{B}^{-1}(\mathbf{y}_1^U)}\prod_{t=1}^T P(a_t|\mathbf{h}_1^T),
\end{equation}
where $a_t \in \mathcal{Y}\cup \{\phi\}$, and $\mathcal{B}$ maps $\mathbf{a}_1^T$ to $\mathbf{y}_1^U$ by removing repeated tokens and $\phi$.


{\bf Conformer-AED:} To test the robustness of early-exits with complex decoders, we use an attention-based encoder-decoder (AED) model~\cite{Chorowski2015AttentionBasedMF}. 
Retaining the architecture of the Conformer-CTC encoder, we replace its linear decoder with four Transformer layers with cross-attention on $\mathbf{h}_1^T$.
This decoder contains two output heads, trained with a CTC loss and a sequence-to-sequence cross-entropy loss respectively~\cite{Kim2016JointCB}. The overall loss function is: 
\begin{equation}
\label{eq:SBloss}
    \mathcal{L}_{\mathrm{AED}}(\hat{\mathbf{y}},\mathbf{y}) = \lambda_{\mathrm{CTC}}\mathcal{L}_{\mathrm{CTC}}(\hat{\mathbf{y}},\mathbf{y}) + \lambda_{\mathrm{CE}}\mathcal{L}_{\mathrm{CE}}(\hat{\mathbf{y}},\mathbf{y}),
\end{equation}
where $\mathcal{L}_{\mathrm{CE}}(\hat{\mathbf{y}},\mathbf{y})=-\sum_{u=1}^U \log P(y_u|\mathbf{h}_1^T,\mathbf{y}_1^{u-1})$, and $\lambda$'s are hyperparameters.
Following the SpeechBrain recipe~\cite{Ravanelli2021}, we set $\lambda_{\mathrm{CTC}}$ and $\lambda_{\mathrm{CE}}$ to 0.3 and 0.7, respectively. 
During inference, only the cross-entropy head is used, 
and a Transformer-based language model trained with the same tokenization is used to rescore the hypothesis.


{\bf Wav2Vec2-CTC and WavLM-CTC:} Whereas the Conformer-based models are trained \textit{from scratch} by optimizing Eq. \ref{eq:jointloss}, for Wav2Vec2-CTC and WavLM-CTC we instead apply early-exit fine-tuning on a \textit{pre-trained} encoder (Wav2Vec-2.0~\cite{baevski2020wav2vec2} and WavLM Base+) encoder 
using the joint CTC losses~(Eq.~\ref{eq:ctc}). The models operate on raw waveforms processed using a convolutional feature extractor, which is frozen during fine-tuning. Note that the pretraining of the two backbones is also different, as WavLM Base+ is trained on $94k$ hours of English speech.


\section{Early-exit selection}
\label{sec:inference}

For an early-exit model at inference time, an uncertainty measure can be used to decide at which exit to output. That is, an exit layer is selected when its uncertainty drops below a given threshold that is estimated to guarantee a desired performance level. Since the encoder layer outputs are converted to posterior probabilities through a softmax module, their uncertainties are suitably represented by their average frame entropies:
\begin{equation}
\Xi^m = -\frac{1}{T|\mathcal{Y}|}\sum_{t=1}^{t=T}\sum_{y\in \mathcal{Y}}P[y|{\bf h}^m_t]\log(P[y|{\bf h}^m_t]) \\
\label{eq:entropy}
\end{equation}

\noindent where, $P[y|{\bf h}^m_t]$ is the probability in the $m^{th}$ encoder output at time $t$  for each output token $y\in\mathcal{Y}$. 
While entropy is a common choice in literature, we also investigate a metric that estimates sentence confidence, computed by applying a softmax to the scores of the N-best hypotheses provided by each decoder:

\begin{equation}
 \Psi^m = \frac{e^{s^m_1}}{\sum_{1}^{K} e^{s^m_k}}
\label{eq:posterior}
\end{equation}
where $s^m_k$ is the log-probability of the $k^{th}$ hypothesis at layer $m$, i.e. $s^m_k=\log(P[\hat{{\bf y}}^m_k|{\bf X};{\bf \Theta}_m])$, and $K$ is the number of N-best hypotheses. Preliminary experiments, aimed at finding the optimal performance/complexity trade-off, suggested the  value $K=300$. 

\section{Experiments}

We carry out experiments using LibriSpeech~\cite{Libri}, TED-LIUM (release 3, \cite{tedlium}), and VoxPopuli \cite{wang-etal-2021-voxpopuli}. LibriSpeech contains $\approx$1,000 hours of read-aloud audiobooks, partitioned into $\approx$960h for training and $\approx$20h for evaluation. 
TED-LIUM comprises of $\approx$452h of transcribed English speeches from TED video conferences for training and $\approx$6h for evaluation. VoxPopuli is a multi-lingual corpus formed of $\approx$400K hours of recordings from European Parliament events. For this work, we use the English subset, consisting of $\approx$543h for training $\approx$60h for evaluation. 

\subsection{Implementation details}

We consider four models: Conformer-CTC (Eq.~\ref{eq:ctc}), Conformer-AED (Eq.~\ref{eq:SBloss}), Wav2Vec2-CTC and WavLM-CTC. 
The Conformer models take as input $80$ Mel Frequency Cepstral Coefficients (MFCCs). 
This MFCC sequence is passed through 
a series of 1D convolution sub-sampling layers. The output of this block 
is applied to a positional encoding module that feeds a stack of 12 Conformer blocks. Wav2Vec2-CTC also consists of a convolutional feature extractor followed by a 12-layer self-attention encoder, but instead takes raw waveforms as input. 
%
Both Conformer-CTC and Conformer-AED use a BPE-based tokenizer~\cite{bpesennrich2016}, with 256 and 5000 tokens respectively. The exit decoders of Wav2Vec2-CTC instead produce 32 grapheme-based tokens (28 characters + 1 blank token + 2 sentence boundary tokens + 1 unknown token) per its official recipe.


The code for both Conformer-CTC and Wav2Vec2-CTC models is available\footnote{\url{https://github.com/augustgw/early-exit-transformer} and \url{https://github.com/augustgw/wav2vec2-ee}}, while the Conformer-AED model and WavLM-CTC follow the related SpeechBrain recipe. Tab.~\ref{tab:hyperparams} summarises the main hyperparameters for the four models.


\begin{table*}[tb]
\centering
\caption{
WERs of the 4 models on LibriSpeech at different layers. “EE”: early-exit models; “no-EE”: individual single-exit models trained independently. “Layer”: exited layer in EE models or total number of layers in no-EE models. Due to time constraints, experiments with No-EE models are focused on low layers, which are the most relevant, and are not available for all cases ("-").
}
\label{tab:results_scratch_libri}
\centerline{
\begin{tabular}{@{}c@{\hskip 3ex}rrrr@{\hskip 3ex}rrrr@{\hskip 3ex}rrrr@{\hskip 3ex}rrrr@{}}
\toprule
\multirow{3}{*}{\textbf{Layer}} &\multicolumn{4}{@{\hskip 3ex}c@{\hskip 3ex}}{\textbf{Conformer-CTC}}&\multicolumn{4}{@{\hskip 3ex}c@{\hskip 3ex}}{\textbf{Conformer-AED}}&\multicolumn{4}{@{\hskip 3ex}c@{\hskip 3ex}}{\textbf{Wav2Vec2-CTC}}&\multicolumn{4}{@{\hskip 3ex}c@{\hskip 3ex}}{\textbf{WavLM-CTC}}\\
\cmidrule(l{1ex}r{1ex}){2-5} \cmidrule(l{1ex}r{1ex}){6-9} \cmidrule(l{1ex}r{1ex}){10-13} \cmidrule(l{1ex}r{1ex}){14-17}

& \multicolumn{2}{c}{\texttt{test-clean}} &  \multicolumn{2}{c}{\texttt{test-other}} &  \multicolumn{2}{c}{\texttt{test-clean}} &  \multicolumn{2}{c}{\texttt{test-other}} &  \multicolumn{2}{c}{\texttt{test-clean}} &  \multicolumn{2}{c}{\texttt{test-other}} &  \multicolumn{2}{c}{\texttt{test-clean}} &  \multicolumn{2}{c}{\texttt{test-other}} \\
\cmidrule(l{1ex}r{1ex}){2-3} \cmidrule(l{1ex}r{1ex}){4-5} \cmidrule(l{1ex}r{1ex}){6-7} \cmidrule(l{1ex}r{1ex}){8-9} \cmidrule(l{1ex}r{1ex}){10-11} \cmidrule(l{1ex}r{1ex}){12-13} \cmidrule(l{1ex}r{1ex}){14-15} \cmidrule(l{1ex}r{1ex}){16-17}
    & no-EE & EE & no-EE & EE & no-EE & EE & no-EE & EE& no-EE & EE& no-EE & EE & no-EE & EE & no-EE & EE \\
    \midrule
  2 & 17.6 & 23.9 & 36.1 & 43.8 & 18.9 & 20.1 & 38.0 & 40.1& 35.7 & 33.7 & 56.7 & 56.0 &  28.2 &  28.0 &  47.6 &  48.5 \\
 4 &  9.8 & 11.6 & 24.3 & 25.7  & 12.8 & 12.5 & 25.8 & 25.2&  17.4 & 17.4 &  35.5 & 36.7 &  12.9 &  13.9 &  27.2 &  27.3 \\
6 &  7.6 &  6.8 & 20.0 & 18.1& 8.4 & 7.7 & 20.1 & 17.1 & 10.7 & 9.6 & 24.8 & 23.7 &  -- &  8.7 &  -- &  18.4 \\
8 &    -- &  5.9 &    -- & 16.3  & -- & 4.4 & -- & 11.5&  -- &  5.8 &  -- & 15.9 &  -- &  4.8 &  -- &  12.4 \\
10 &    -- &  5.2 &   -- & 15.8  & -- & 2.8 & -- & 6.9&  -- & 4.5 &   -- & 12.6 &  -- &  4.0 &  -- &  9.5 \\
12 &  6.5 &  5.1 & 17.7& 15.1 & 2.5 & 2.3 & 6.1 & 6.0 & 3.4 & 4.3 & 8.6 & 12.2 &  3.0 &  3.6 &  6.5 &  8.8 \\
\bottomrule 
\end{tabular}
}
\end{table*}

\section{Results}
\label{sec:results}  

All results reported in this section are expressed in terms of word error rates (WERs) computed on the standard test partitions of the three datasets.  
Tab.~\ref{tab:results_scratch_libri} reports the 
performance on LibriSpeech at different exits, training Conformer-CTC and Conformer-AED models from scratch and fine-tuning Wav2Vec2-CTC and WavLM-CTC. For each model, we also report the performance of the corresponding single-exit model for comparison. 

The Conformer-CTC model with 12 layers achieves 6.5\% on test-clean and 17.7\% on test-other. As expected, WER is higher in the lower layers. The performance decreases significantly only in the lowest two exits (Layer 2, 4). In the middle layers (6, 8, 10), performance is comparable to that of the uppermost layer, while requiring significantly fewer parameters. 
Similar trends are achieved with Conformer-AED, but with significantly better overall performance (2.3\% and 6.0\% WER in the 12th layer for test-clean and test-other, respectively). 
This absolute improvement is attributed to the Transformer-based decoders and language model rescoring, allowing to reach state-of-the-art on LibriSpeech. Tab. ~\ref{tab:results_scratch_libri} shows that both models based on pre-trained backbones (Wav2Vec2-CTC and WavLM-CTC) exhibits similar behaviors. However, since in pretraining they have been optimised solely on the loss of the highest layer, they displays higher WERs at lower exits with a noticeable performance gain beginning at the 8th layer. This low-layer degradation is much less evident in the both models trained from scratch. 
In summary, although smaller and trained on less data, the Conformer-CTC/AED models perform better than those based on pre-trained models in the lowest three layers. Apart from the lowest exits, the early-exit Conformer-CTC/AED models achieve better WERs than the corresponding single-exit counterparts (column "no-EE" in Tab. ~\ref{tab:results_scratch_libri}).
This indicates the beneficial effects of the compound loss, acting as a regulariser and improving both robustness and generalization, as observed in previous studies incorporating losses at lower layers 
\cite{scardapane2020, branchynet2017, geng2021romebert}. In other words, using a single model with multiple exits not only reduces the computational burden of training multiple single exit models but also delivers better performance. Note that the same behavior is not observed for both the pre-trained models, where the performance of the 12th layer of the EE model is worse than the corresponding 12-layer no-EE model. These results reflect that Wav2Vec2-CTC and WavLM-CTC have not been trained from scratch but only fine-tuned with early exits.

{\bf These findings suggest that for early-exit architectures, training a model from scratch might be more efficient than fine-tuning a large but conventionally pre-trained model.} It is worth noting that the same trends are observed considering different decoders, different training losses, and independently of the use of a language model. We hypothesise that both Wav2Vec2.0 and WavLM models would display similar improvements in low-layer performance if trained from scratch with early-exits. Unfortunately, these experiments were not affordable.

Finally, experiments on TED-LIUM and VoxPopuli, shown in Tab.~\ref{tab:results_voxted}, confirm the observations drawn on LibriSpeech. In these experiments, we also observe superior performance in models trained with the compound early-exit loss as compared to those trained with single exits, for layers higher than $4$. Moreover, experiments on LibriSpeech-100h (not reported here for the sake of space) have shown similar trends, suggesting the effectiveness of the approach also for low-resource ASR settings.

\begin{table}[t]
\centering
\caption{WERs on the {\bf TED-LIUM} and {\bf VoxPopuli} at different exits, training the model Conformer-CTC  from scratch.
}
\label{tab:results_voxted}
\begin{tabular}{@{}crrrr@{}}
\toprule
\multirow{2}{*}{\textbf{Layer}}
 &\multicolumn{2}{c}{\textbf{TED-LIUM}}&\multicolumn{2}{c}{\textbf{VoxPopuli}} \\
 \cmidrule(r){2-3} \cmidrule(l){4-5}
 & no-EE & EE& no-EE & EE\\
\midrule
2  &42.7 &43.8 &  27.3 &36.7 \\
4  & 35.4   &23.4 &   19.7    &21.1\\
6  &25.5 &18.0 & 18.7 &17.3\\
8  &  --   &16.1 & --  &15.4\\
10 &  --   &14.9 & --  &14.7\\
12 &16.4 &14.6 & 16.3 &14.3\\
\bottomrule
\end{tabular}
\end{table}

\subsection{Exit selection during inference}
\begin{figure}[t]
  \centering
  \centerline{\includegraphics[trim={1.8cm 0 2.3cm 0},clip,width=7cm]{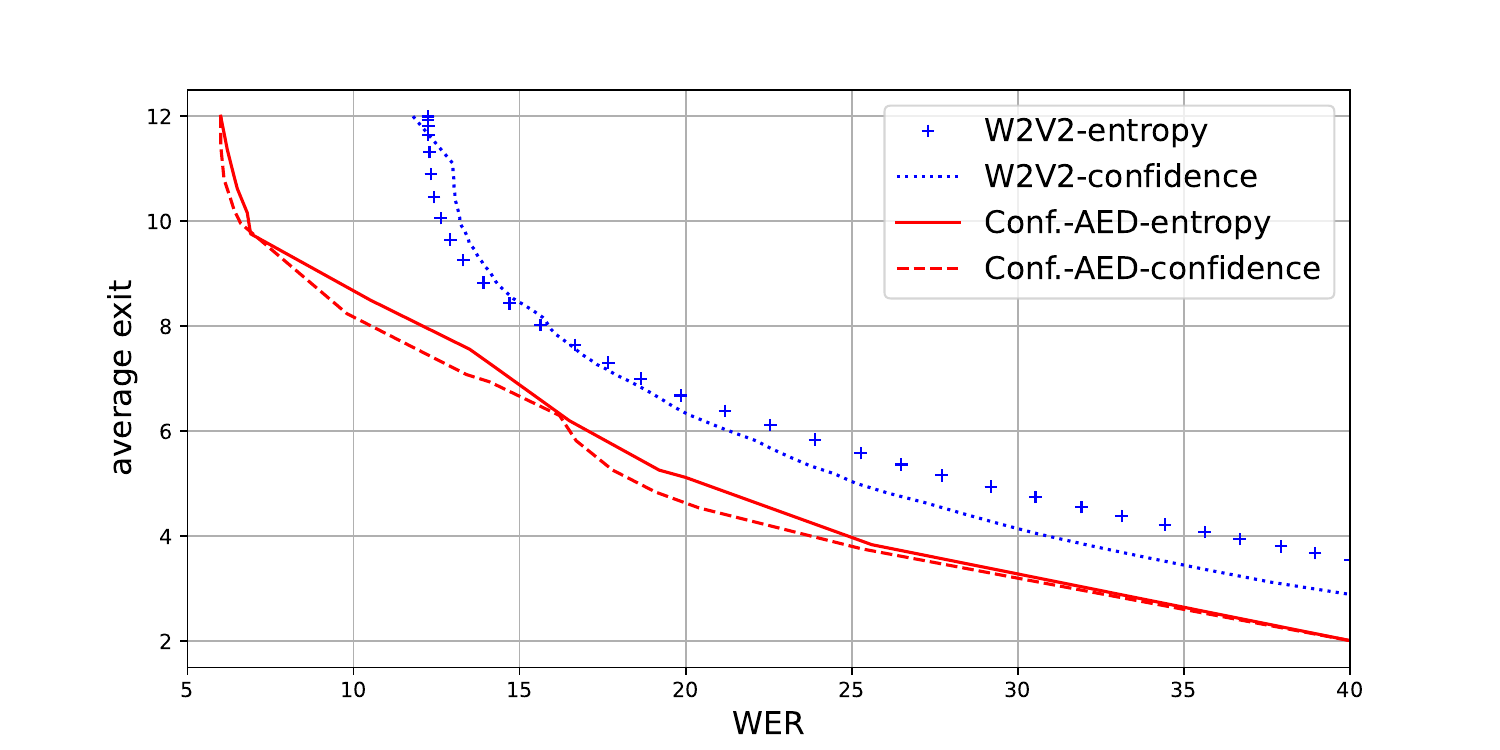}}
\caption{Average exit selection (y-axis) and WER (x-axis) when varying the exit selection threshold for Conformer-AED and Wav2Vec2-CTC, using both entropies and sentence confidence as exit metrics.}
\label{fig:rocinference}
\end{figure}

In this section we analyse exit selection at inference time using either average frame entropy (Eq.~\ref{eq:entropy}) or sentence confidence (Eq.~\ref{eq:posterior}). 
We follow the common thresholding approach, selecting the first exit below a predefined entropy threshold or above a predefined posterior threshold. Previous studies~\cite{zaiem2023,berrebbi2022avoid} have observed that although the overall performance of lower layers is typically inferior to that of the final layers, in many cases the performance is on par. Being able to identify those cases would considerably reduce the overall computational cost.
Fig.~\ref{fig:rocinference} shows the average selected exit with the corresponding WER when varying the threshold for two models and the two metrics (the other models are not reported for the sake of readability). The closer the curve to the chart origin, the better.
We observe that, as expected, better models deliver better performance in exit selection: the Conformer-AED lines are well below the others. Sentence confidence (dotted lines) on average selects lower exits than entropy at the same WER values, indicating that it provides a better trade-off between saving computation and maintaining performance. However, estimating the sentence confidence maybe not always computationally viable and other factors, such as the computational capabilities of the hosting device, may influence which exit selection method is most appropriate.


\section{Conclusion and future works}
In this paper, we have investigated early-exit architectures for ASR by comparing the training and inference of models based on the Conformer and on pre-trained models (Wav2Vec2 and WavLM). In contrast to previous studies, we bring additional attention to the training dynamics of early-exit models. In particular, we demonstrated the benefits of training models from scratch using early exits, as compared to fine-tuning a pre-trained model, on three datasets. Future works will investigate weighting schemes for the compound loss in Eq.~\ref{eq:jointloss} or alternative training strategies~\cite{scardapane2020}, including distillation (similar to \cite{geng2021romebert}) from upper layers. Additionally, methods to approximate confidence-based methods for exit selection could improve their practicality on low-resource devices. 
\\
{\small
\begin{spacing}{0.85}
\noindent
\textbf{Acknowledgments.} This work was partially funded by the PNRR ICSC National Research Centre for High Performance Computing, Big Data and Quantum Computing (CN00000013), and by PNRR project FAIR - Future AI Research (PE00000013) under the NRRP MUR program funded by the NextGenerationEU. We also acknowledge support from the JSALT23 workshop, hosted at Le Mans University, France, and sponsored by Johns Hopkins University with unrestricted gifts from Amazon, Facebook, Google, and Microsoft.
\end{spacing}
}


\begin{small}

\bibliographystyle{IEEEbib-abbrev}
\bibliography{strings,refs}
\end{small}
\end{document}